# Multisource Adaptive Data Distribution and Routing in Wireless Sensor Networks

**Subhabrata Mukherjee[1], Amrita Saha[1] Mrinal K. Naskar[2] and Amitava Mukherjee[3]**

[1]Dept of CSE, IIT Bombay, Mumbai 400 076, India

[2]Advanced Digital and Embedded Systems Lab, Dept of ETCE, Jadavpur University, Calcutta 700 032, India

[3] IBM India Pvt Ltd, Salt Lake, Calcutta 700 091, India

amitava.mukherjee@in.ibm.com, subhabratam@cse.iitb.ac.in , amrita@cse.iitb.ac.in

**Abstract** – The wireless sensor network is a collection of energy-constrained nodes. Their objective is to sense, collect and process information for some ad-hoc purpose. Typically the nodes are deployed in geographically inaccessible regions. Thus the most challenging task is to design a network with minimal power consumption. As the nodes have to collect and process data very fast, minimizing data delivery time is another objective. In addition to this, when multiple sources transmit data simultaneously, the network load gradually increases and it may lead to congestion. In this paper we have proposed an adaptive framework in which multiple sources transmit data simultaneously with minimal end-to-end data delivery time and minimal energy consumption besides ensuring that congestion remains at an optimum low so that minimal number of data packets are dropped. This paper presents an adaptive framework to achieve the above-mentioned objectives.
This framework has been used over Mac 802.11 and extensive simulations have been carried out in NS2 to prove the effectiveness of the framework over traditional Mac as well as few other existing protocols.

## 1. INTRODUCTION

The static sensor network [1]-[3] consists of a large number of smart sensor nodes distributed randomly in a geographically inaccessible area. The minimization of power consumption in WSN is one of the most important design issues because sensor nodes are deployed mostly in geographically inaccessible areas and their energy content thus cannot be easily replenished. However, there are cases like surveillance [1] in battlefield, where the movement of enemy troops is to be monitored continuously. In these cases data must be sensed, processed and transmitted very quickly, so multi-path routing scheme can be used to reduce the data delivery time. Earlier researches [1], [2], [4], [5] showed that multi-path communication can improve the end-to-end data delivery time by taking recourse to simultaneous data transfer over multiple spatial paths.

In this work, we have proposed a method to divide the data into blocks, taking various network factors into consideration, and sending them simultaneously along the different paths available. We have implemented this framework with Mac 802.11 and have shown that it achieves an optimal data delivery time and power consumption when compared to a few other existing protocols as well as the traditional Mac 802.11. We first propose a framework that considers the network to be a single event model in which only one event can be served at a time i.e. only one source can transmit at a time while others have to wait for their turn. We then propose a multi-event framework where multiple sources can transmit simultaneously which is obtained by the superposition of multiple single event models (each corresponding to a single source). An effective strategy for data distribution and scheduling the dispatch of the data packets has also been discussed.

The organization of the rest of the paper is as follows: Section 2 gives a related work section. Section 3 describes the system framework for a single source model. Section 4 enhances this framework to introduce an adaptive mechanism for a single source multi-path routing and strategic data distribution along the multiple paths to minimize data delivery time and the net energy consumption. Section 5 extends the single-source model to suit multiple sources and outlines an effective strategy for handling and dispatching of data packets to minimize the data delivery time, net energy consumption as well as congestion in the network. Section 6 gives the simulation results for the single-source and the multi-source model followed by concluding remarks.

## 2. RELATED WORK

There has been a lot of research works in the wireless sensor network area. There have been extensive studies on routing and data distribution in wireless sensor networks. A number of metrics have been used to assess the routing quality, among which the most common and widely used metric has been the hop count. The protocols that use shortest path routing include Dynamic Source Routing (DSR) [6][7], Ad-hoc On-demand Distance Vector routing (AODV) [8], Destination Sequence Distance Vector Routing (DSDV) [9]. In this paper we have taken this as Scheme 1 where all the data has been forwarded via the path with the minimum hop count from the source to the destination. Network reliability can be improved by using multiple paths from the source to the destination instead of using a single path [1][4][5]. In [10], multipath routing is used to increase the reliability of WSNs. The proposed scheme splits up the data into smaller sub packets of equal size and sends them via the multiple paths available. We take this as Scheme 2 where all the data is split up equally and sent across the multiple spatial paths available. We have shown that our first proposed framework (Scheme 3) is better than the above 2 schemes as our suggested scheme takes various factors like hop count, energy dissipation due to transmission and reception, bit rate and various other network factors into consideration while distributing the data along multiple spatial paths. We have extended our first framework for a single source to suit multiple sources. We refer to the extended suggested framework as the MADDR (Multisource Adaptive Data Distribution And Routing). [11] suggests an efficient multipath protocol (DCHT) for the wireless sensor network and establishes its efficiency over some other protocols like the Directed Diffusion [12], EDGE [13], C-MFR [14]. We have shown by simulation that the proposed MADDR when implemented with Mac 802.11 gives a better throughput than DCHT over different network sizes. Furthermore, we have compared the data delivery time and the net energy consumption in MADDR over MAC 802.11 with a few other existing routing protocols like the AODV, DSR and DSDV on Mac 802.11 in NS2. Our proposed MADDR framework achieved a better performance in terms of data delivery time, net energy consumption and less dropped packets due to congestion.

## 3. SYSTEM FRAMEWORK FOR A SINGLE-SOURCE MODEL

### System Description

We consider a large number of smart sensors randomly distributed in a geographically hostile area. Any sensor can act as a source node and there is only one sink node (base station).
The sink node alone will receive all the data sent by sensors. Each sensor operates on limited battery. Energy is consumed mostly in transmission and receiving data at its radio transceiver. Energy is also consumed when the nodes are sensing or processing data. Each node senses information and delivers it to the sink through a set of paths, each comprising of multiple hops. Each sensor $n \in N$ has a unique identifier. The data sensed by each node is divided among the spatial multi-paths so that the energy consumed in the process and the net end-to-end data delivery time is minimized.

### 3.2 Communication Delay

The data delivery time in any path while sending data from the source to the sink consists of two components:

a) *Queuing/processing delay*: we consider $q_j$ as the average queuing delay per packet per hop for the jth path.

b) *Transmission/reception delay*: this delay per packet per hop for the jth path is modeled as $S/b_j + l_j$.

(where S is the packet size, $b_j$ is the link speed in bits/sec in the jth path and $l_j$ is the link delay in the jth pth).

Thus the amount of time required for a data packet to traverse a link (over one hop) along with the queuing delay is defined to be

$$\tau_j = s/b_j + d_j + q_j \tag{1}$$

Thus, the total delay in the jth path to send $\Delta_j$ data packets over $H_j$ hops is given by

$$T_j = (\Delta_j * \tau_j * H_j) \tag{2}$$

### 3.3 Energy Consumption

Energy consumption in WSN can be largely categorized into two parts:
  a) *Communication*
  b) *Sensing and processing*

The communication related energy consumption is due to transmission and reception. First we find an expression for the total energy consumed by all the nodes in the jth path.
An energy dissipation model for radio communication similar to [15], [16] has been assumed. As a result the energy required per second for successful transmission for each node ($E_{ts}$) is thus given by,

$$E_{ts} = e_t + e_d d^k \tag{3}$$

(where $e_t$ is the energy dissipated in the transmitter electronic circuitry per second to transmit data packets and $e_d d^k$ is the amount of energy required per second to transmit over a distance d and k is the path loss exponent (usually $2.0 \leq k \leq 4.0$)).

The distance d, must be less than or equal to the radio range $R_{radio}$, which is the maximum inter-nodal distance for successful communication between two nodes. If $T_{1b}$ is the time required to successfully transmit a bit over a distance d then total energy to transmit a bit for each node is

$$E_t = (e_t + e_d d^k) T_{1b} \tag{4}$$

If $e_r$ is the energy required per second for successful reception and if $T_{2b}$ is the total time required by a sensor to receive a bit then the total energy to receive a bit for each node is

$$E_r = e_r T_{2b} \tag{5}$$

If we take $d_j$ as the *average inter-hop* distance in the jth path,
then $d_j$ can be approximated as $d_j = T/H_j$
(where T is the total distance between source and sink and $H_j$ is the number of hops in the jth path).

If $\Delta_j$ is the number of packets pushed in the jth path, the total energy dissipation due to communication by each node in the jth path is $(E_t + E_r) * \Delta_j * S$.
If $H_j$ is the number of hops in the jth path, the number of nodes in the jth path is given by $H_j + 1$.
Power dissipation at each node due to minimum computation and sensing can be assumed to take place approximately at an effective rate $K_r$.
The total power consumed by all the nodes due to sensing and processing is equal to $K_r *$(number of nodes) which is independent of the data division.

Thus the total energy dissipation in the jth path is
$$E_j = [E_t + E_r] * \Delta_j * S * (H_j + 1) + K_r * (H_j + 1)$$
$$= [(e_t + e_d *(T/H_j)^k) * T_{1b} + e_r * T_{2b}] * \Delta_j * S * (H_j + 1) + K_r * (H_j + 1) \qquad (6)$$

## 4 ADAPTIVE MECHANISM FOR MULTI-PATH ROUTING FOR A SINGLE-SOURCE MODEL

### 4.1.1 Creation of the routing table and variable estimation for a Single Source Model

When a sensor node first joins the network it finds a set of paths to each of the other nodes in the such that the paths are mutually node exclusive i.e., the nodes in the jth path are distinct from those of the ith path
  During route discovery, the source node broadcasts "hello" packets, which contain the id of the source node and special control information so that other nodes can identify that packet as a special control packet. Each node on receiving a "hello" packet sends a "reply" packet as soon as it can back to the source node which contains its id and other parameters like $K_r$, $H_j$, $e_t$, $e_d$, $T_{1b}$, $T_{2b}$. To get $\tau_j$ of each route, the source node divides the total time between the sending of the "hello" packet and getting a "reply" packet from each node by two. Note that, using hello packets we do not require the values of the different components of $\tau_j$ i.e., $b_j$, $d_j$ and $q_j$.
  Each node creates a routing table containing the above-mentioned information. Each entry is indexed by a destination node and a set of paths (having the above mentioned characteristics) to reach the destination node and information about those paths as obtained during the route discovery phase.
  The sensing event in WSN is assumed to be event driven As soon as a sensor detects an event in its vicinity it checks its routing table. By using multipath it distributes data among the paths obtained from the routing table. The data (D) sensed by a sensor node is thus divided into datasets $\Delta_j$ for j ∈ [1, n], which is distributed over multiple spatial paths which is done in such a way so as to optimize the end-to-end data delivery time and the power consumption of the network. *The creation of routing table and updation presents an overhead but this is done only once during network setup or when there are multiple node/link failure or multiple new nodes come up.*

### 4.1.2 Three Schemes and Optimization Algorithm

  Earlier researches [1], [2], [4], [5] showed that if we send all the data packets from source to sink via a single path (*Scheme 1*) then the end-to-end data delivery time is often more than that measured when we distribute the data packets equally in all paths and send them simultaneously (*Scheme 2*).
  In ideal situations with no congestion, no retransmission and transmission power of all the nodes being the same in both single and multiple paths, the total power consumption in Scheme 2 is often greater than that in Scheme 1 due to the involvement of more number of sensors in the multiple paths, route discovery mechanisms involving a greater number of sensor nodes, greater cumulative power dissipation due to sensing, tranmission, reception etc.
  Our objective is to distribute and route the data packets through several paths available from the routing table for each source node in such a way that the end-to-end data delivery time is even less than *Scheme 2* and the total power consumption of the system lies between that of *Scheme 1* and *Scheme 2* but close to that of Scheme *1*. We call this *Scheme 3*.
  If we analyse the expression term of $E_j$ we find that $E_j$ decreases as $H_j$ increases which means short hops are favourable. But on the other hand as $H_j$ increases delay also increases due to processing by multiple nodes. Hence we need a trade-off between delay and energy consumption.
  We introduce a term here called the energy-delay product (EDP).
The EDP for the jth path is defined as,
$$EDP_j = E_j * T_j$$
$$= [\{(e_t + e_d *(T/H_j)^k) * T_{1b} + e_r * T_{2b}\} * \Delta_j * S * (H_j + 1) + K_r * (H_j + 1)] * (\Delta_j * \tau_j * H_j) \qquad (7)$$

Now to reduce the overall data delivery time and the net energy consumption of *Scheme 3* , we seek to make the energy-delay product of every path of Scheme 3 less than or equal to the average energy-delay product of Scheme 2.

The average energy-delay product ($EDP_{avg}$) of Scheme 2 is given by,

$$EDP_{avg} = E_{avg} * T_{avg}$$
$$\approx [ \{( e_t+e_d *(T/ H_{avg})^k)* T_{1b} + e_r *T_{2b} \}*(D/n)*S*(H_{avg}+1)+K_r*(H_{avg}+1) ] * ((D/n)*\tau_{avg}*H_{avg}) \quad (8)$$

($\Delta_j=D/n$, as data is distributed equally in all the paths in Scheme 2).
Thus to make the net energy-delay product of Scheme 3 less than that of Scheme 2, $EDP_j <= EDP_{avg}$ for j$\in$ [1, n]. *Hence*,

$$[ \{( e_t+e_d *(T/ H_j)^k)* T_{1b} + e_r *T_{2b} \}*\Delta_j*S*(H_j +1)+K_r*(H_j+1) ] * (\Delta_j*\tau_j*H_j) <= [ \{( e_t+e_d *(T/ H_{avg})^k)* T_{1b} + e_r *T_{2b} \}*(D/n)*S*(H_{avg}+1)+K_r*(H_{avg}+1) ] * ((D/n)*\tau_{avg}*H_{avg}) \quad (9)$$

Subject to the constraint,

$$\sum_j \Delta_j = D \text{ for } j \in [1, n] \quad (10)$$

Equation (9) gives the maximum number of data packets $\Delta_j$ that can be pushed in the jth path keeping the data delivery time and the net energy consumption less than that of *Scheme 2*.

Here one may argue that by reducing the EDP of the jth path it may so happen that one component of the EDP may increase and the other may decrease so that the overall EDP for the jth path decreases. So our objective of reducing both the components of EDP i.e data delivery time and net energy consumption may not be achieved.

But this argument does not hold water since the *only variable for the jth path is* $\Delta_j$. The equation above gives a threshold value of $\Delta_j$ that can be pushed in the jth path keeping both energy consumption and delay under a threshold value. Since the data delivery time and the net energy consumption are both an increasing function of $\Delta_j$, for any path if one of them decreases the other is also bound to decrease depending on $\Delta_j$, keeping all other factors constant.

The solution to (9) + (10) will give the value of $\Delta_j$ for j$\in$ [1, n] i.e., the number of data packets to be sent in the jth path in *Scheme 3*. Generally, the computing resources at a node are limited and the classical optimization problem solving techniques require significant computational resource and time. Here we develop a method that does not require much computational resource or time although it may give a sub optimal solution, which nevertheless achieves our purpose.

Equation (9) is of the form $A\Delta_j^2 + B\Delta_j \leq C$ (where A, B, C are constants). We want to find the maximum value of $\Delta j$ satisfying the condition above. So, we first solve equation (9) replacing the inequality sign by equality i.e.,

$$[\{(e_t+e_d *(T/ H_j)^k)* T_{1b} +e_r*T_{2b} \}*\Delta_j*S*(H_j +1)+K_r* (H_j+1) ]*(\Delta_j*\tau_j*H_j) =[\{(e_t+e_d *(T/H_{avg})^k) *T_{1b}+e_r*T_{2b}\} *(D/n)*S*(H_{avg}+1) +K_r*(H_{avg}+1) ] * ((D/n)*\tau_{avg}*H_{avg}) \quad (11)$$

The R.H.S of the equation is a constant as the values of all the terms there are obtained from the routing table. The L.H.S of the equation is only a function of $\Delta_j$ and $H_j$ for j $\in$ [1, n].

The above equation (11) is easily solved by substituting values of the constants and the value of $H_j$ *for each of the n spatial paths*.

The data volume $\Delta_j$ for j $\in$ [1, n] obtained from the equation (11) is the maximum number of data packets that can be sent in the particular jth spatial path.

To satisfy constraint (10), the actual number of data packets to be sent into each jth spatial path j $\in$ [1, n] is given by

$$(\Delta_j /\sum_j \Delta_j)*D \quad for j \in [1, n] \quad (12)$$

### 4.2 Detection of faulty nodes or links

If there is any packet drop it can be due to collision, node or link failure. In case of a packet drop it needs to be retransmitted. If the 2 sources keep on colliding with each other and the data packets are dropped again and again, then after a maximum number of, say, 'm' attempts (typically the value of 'm' is taken as 10) the process is aborted.

Now, either the transmitting node 'a' or the receiving node 'b' or the link connecting 'a' or 'b' is at fault. In that case 'a' performs a self-check by transmitting a beacon packet to another neighboring node 'c'.

Case 1: If 'a' fails in this case too then it is at fault. Every node starts a timer from the time it is supposed to receive any data packet from its neighboring node in any transfer round. In case 'b' fails to receive any packet from 'a' within m* $\tau_j$ secs, then it concludes 'a' is at fault.

Case 2: If 'a' succeeds to send the beacon packet to 'c', then either 'b' or the link connecting 'a' and 'b' is at fault.

Node 'b' in Case (1) and node 'a' in Case (2) chooses the redundant node 'd' nearest to it and assigns it the serial number of the node that just failed. The information is communicated to the neighboring nodes of the failed node so that they can update their routing tables. Then the communication proceeds normally. For the remaining of the data transfer round and till the failed node is repaired, the node 'd' performs the functions of node 'a' in Case (1) and node 'b' in Case 2.

### 4.3 Route discovery Algorithm: some important features

a) The source node only initiates route discovery algorithm when it joins the network or there is any change in the network due to failure of multiple nodes or when multiple new nodes join.
b) The paths discovered are all node disjoint.
c) The route discovery algorithm gives a set of paths between source and sink along with the information about various node and path parameters in each jth path like the values of $K_r$, $H_j$, $e_t$, $e_d$, $T_{1b}$, $T_{2b}$, $\tau_j$

### 4.4 Step by step procedure

The events considered here are non-overlapping in time. In case multiple events occur simultaneously near the vicinity of a node, the events are processed sequentially. Hence while one event is being active, the others have to wait.

*Step 0:* An event occurs near the vicinity of a node. For each event, the following steps are carried out.
*Step 1:* Initialize the set P=Φ, where P is the set of paths for that event in a spatial domain.
*Step 2: If the routing table is not created or there is any change in the network since the routing table was updated, then* call the route discovery algorithm to determine
a) Node-disjoint multi-paths. The maximum number of multi-paths will be less than or equal to the number of nodes within the transmission range of the source node as we consider only node disjoint multi-paths
b) Parameters $K_r$, $H_j$, $e_t$, $e_d$, $T_{1b}$, $T_{2b}$, $\tau_j$ for each jth path
c) Find the average hop count $H_{avg}$.
d) In a certain period of time (in this paper, 'certain period of time' or 'an interval' will mean the time required for successful transmission of data sensed by a particular sensor to the sink) the energy of each node of the network taking part in transmission/reception decreases by $(E_t + E_r) * \Delta_j * S + Kr$ in the jth path. The power of any node not taking part in transmission will decrease by $K_r$ where $K_r$ denotes the effective rate (statistically averaged) of power loss of each node due to sensing.
   *Else* consult the routing table to obtain the above mentioned information.
*Step 3:* Solve equation (9) + (10) with the set of input values determined by the route discovery algorithm
*Step 4:* End

### 5. ADAPTIVE MECHANISM FOR MULTI-PATH ROUTING FOR A MULTI-SOURCE MODEL

#### 5.1 Creation of the routing table and variable estimation for a Multi Source Model

In the multi-source framework each source node acts independently and executes all the steps formulated in section 4.4 regardless of any source node in the network. Thus the set of paths used by each source will be *locally* node-disjoint to it but may not be (*globally*) node-disjoint to any other source node.

During route discovery, the control packet collects all the information outlined in section 4.1.1. In addition, during every data distribution phase, the source node sends out a "choke" packet, in every node-disjoint path available to it, which is a special control packet that visits each node in a path and checks its queue to find out what fraction of the queue is full. This serves as an important heuristics to estimate the probable nodes, which may be affected by congestion in the network. The control packet keeps a counter and increments it for every node it encounters whose queue is more than (say) 50% full. The control packet returns the value of this counter, (say) $C_j$ to the source node. This value gives the number of probable nodes whose queue may overflow due to increased network load leading to the drop of data packets and thus serves as an indication of congestion in the network.

During route discovery and creation of the routing table [explained in section 4.1.1], each node $X_i$ counts the number of its immediate neighbors (say $N_i$) i.e. those nodes, which come within its direct transmission range. If the total size of the queue allotted to node $X_i$ is $M_i$, then $X_i$ fragments its queue into $N_i$ equal parts, each part being solely allotted to hold the data packets for a particular neighboring node. This implies that each fragmented queue will hold the data packets for a particular neighboring node where they are to be delivered and cannot hold any data packet destined for any other node. Each fragmented queue can hold at most $M_i/(N_i * S)$ data packets (where S is the size of each data packet). Also each node keeps a part of its queue separate for control packets exclusively. This is to prevent the drop of any urgent control/message packets, which may occur when the data queue of a node overflows.

Each packet has a field for precedence or priority. This field is used to ensure that during heavy network load, when the queue of a node overflows, lower priority packets are dropped and not the higher priority ones. Control or urgent message packets are given the maximum priority. In the TCP packets there is a field for 'urg' pointer, which is also set. Each node that receives such a packet has to service it first, suspending all its ongoing work temporarily.

### 5.2 Route discovery Algorithm: some important features

a) Every source node executes the route discovery algorithm in a similar way as outlined in Section 4.3 regardless of the presence of any other source node in the network.
b) Thus the set of routes used by any source will be locally node-disjoint to it but may not be globally node-disjoint i.e. they may overlap with the route of some other source.

### 5.3 Data Distribution in each path

Similar to the route discovery phase, during data distribution too each source acts regardless of the presence of other sources in the network and distributes data in each of its path in a similar fashion as outlined in Section 4.1.2. But there is a slight modification to eqn [12]. The value of $\Delta_j$ given by eqn [12] gives the maximum number of data packets that can be pushed in the jth path keeping the data delivery time and the net energy consumption less than some threshold value in a single source framework. But in a multi-source framework, we have to take in consideration the contention nodes (nodes which may possibly be affected by congestion) when multiple sources are transmitting simultaneously. Thus the number of data packets $\Delta_j$ pushed in the jth path should be inversely proportional to the number of contention nodes in that path. In section 5.1 we have outlined a strategy to estimate the number of probable contention nodes $C_j$ in the jth path used by any source. Thus $\Delta_j$ in eqn [12] is replaced by,

$$\Delta_j = (\Delta_j * (1 - C_j/(H_j+1)) / \sum_j \Delta_j * (1 - C_j/(H_j+1)) ) * D \quad for\ j \in [1, n] \qquad (13)$$

Since $H_j+1$ gives the number of nodes in the jth path used by any source, $C_j/(H_j+1)$ gives the fraction of the total number of nodes in the jth path that may be affected due to congestion and where a probable packet loss may occur.
When a single source is transmitting or when multiple sources are starting together at the same instant at T0 or when the queue of each node in the jth path of a source is less than 50% full, $C_j = 0$ and thus eqn [13] reduces to eqn [12].

### 5.4 Scheduling the dispatch of data packets

We have already outlined the requirement for queue fragmentation in the previous sub-section. If there are $N_i$ fragmented queues in a node $X_i$, then the node will dispatch data packets from each of the $N_i$ queues in a round-robin fashion i.e. first it will dispatch a data packet from queue 1, then from queue 2 and so on and finally from queue $N_i$.

Congestion generally starts at a particular node or link and then spreads throughout the network. When there is a single queue, then packets for all different destination nodes are queued up together in it. Thus, if any particular link begins to get congested and there are a number of consecutive data packets in the queue to be forwarded in that link, this will invariably increase the congestion in that link. Consequently, the data packets, waiting in the queue, that are to be forwarded in other links (which may be relatively less congested) are also delayed as the queue works in a FIFO order. Thus, unless the previous packets are dispatched along the congested-link the latter packets cannot be dispatched. This is the problem of maintaining a single queue. Maintaining multiple queues ensure that the load or congestion in one link does not affect the data distribution in other links which do not face such problems. Furthermore, this scheme also gives the congested link a time quantum, for its congestion to relax a bit, before the next packet is pushed in it. This time quantum is given by the interval when one packet is pushed in that link and when its turn comes again in the round-robin order. Thus during congestion, the load is distributed and no link is affected by the traffic in any other link. Also when there is no congestion, the load still remains distributed reducing the possibility of any particular link getting over congested due to multiple data packets pushed in it from any particular node continuously. These are the principal factors that led to the introduction of queue segmentation and round robin scheduling in the MADDR framework.

### 5.5 GRAPHICAL EXPLANATION OF THE MADDR FRAMEWORK

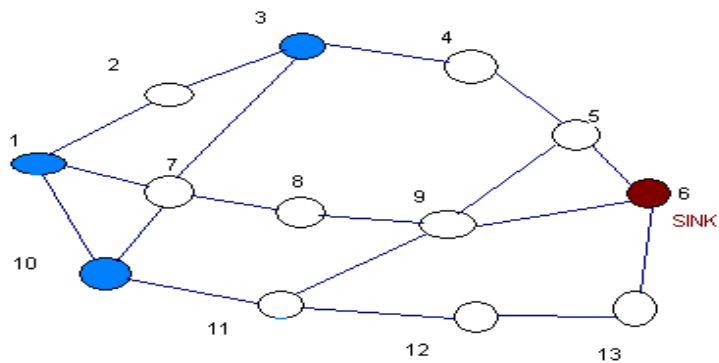

The above diagram shows a network with 13 sensor nodes. The nodes marked in blue are the 3 sources – node 1 is S1, node 3 is S2, node 10 is S3 and node 6 marked in red is the common sink. The 3 *locally node-disjoint* paths for each source are :

For **S1**: Path1 (P11) 1-2-3-4-5-6   ; Path2 (P12) 1-7-8-9-6   ; Path3 (P13) 1-10-11-12-13-6
For **S2**: Path1 (P21) 3-4-5-6        ; Path2 (P22) 3-7-8-9-6   ; Path3 (P23) 3-2-1-10-11-12-13-6
For **S3**: Path1 (P31) 10-11-12-13-6 ; Path2 (P32) 10-7-8-9-6  ; Path3 (P33) 10-1-2-3-4-5-6

Suppose each source wants to send 100 data packets to the sink. If all the 3 sources start transmitting together at the same time at instant T0, the queue of each node will be empty and hence there will be no contention node. Thus the value of $C_j$ for every path, for every source is 0.

After the route discovery phase and variable estimation, each source node performs data distribution, independent of the other source nodes, and computes the number of data packets $\Delta_i$ to be pushed in the jth path using eqn [11]+[12]. Putting the value of the network parameters in eqn [11]+[12] (the value of the network parameters is obtained from Table 1 in Section 6.1), the value of $\Delta_j$ for each source is obtained as:

For S1 : P11 – 30 packets ; P12 – 40 packets ; P13 – packets  30
For S2 : P21 – 45 packets ; P22 – 35 packets ; P23 – packets  20
For S3 : P31 – 37 packets ; P32 – 37 packets ; P33 – packets  26

Now, each node will have its queue fragmented into as many parts as the number of neighbors within its direct transmission range and each part will solely be dedicated to hold the data packets destined for a particular neighboring node. For example, node 3 will have 3 queues, each to hold data packets for nodes 2,4 and 7 respectively, node 9 will have 4 queues corresponding to the nodes 5,6,8 and 11 respectively. Each node will service its queues in a round-robin order as explained before.

There is a constraint to the number of queues that can be formed at a time by fragmenting the original queue. This is dictated by the node's memory storage area, size of each data packet and the number of its immediate neighbors.

## 5  SIMULATION RESULTS

### 5.1  Simulation results for the Single-Source model

In order to study the performance of our proposed framework we have run simulation program in NS2. We have first run the route discovery algorithm. We consider an area of 501*501 square meters where 1000 nodes are deployed randomly. The nodes have a transmission radius of 2.4 meters. The simulation has been done with 100 and 200 data packets (size of each data packet is 1 kb). The simulation has been done for multiple runs and the mean result for each scheme has been shown. In this simulation, the routing algorithm gives 5 possible paths between the source node and the sink along with the following parameters-

| | |
|---|---|
| *Initial energy of each node* | 23760 joules for all (calculated on basis of MICA 2 motes) |
| *Mac* | Mac/802.11 |
| *Bit rate* | 50 kbps |
| $K_r$ | .024 Watts |
| *Transmission Power* per packet per hop | 1024 µJ/sec |
| *Receiving Power* per packet per hop | 819.2 µJ/sec |
| *Idle Power* | 409.6µJ/s |

**Table 1 : Network Parameters used in the Simulation**

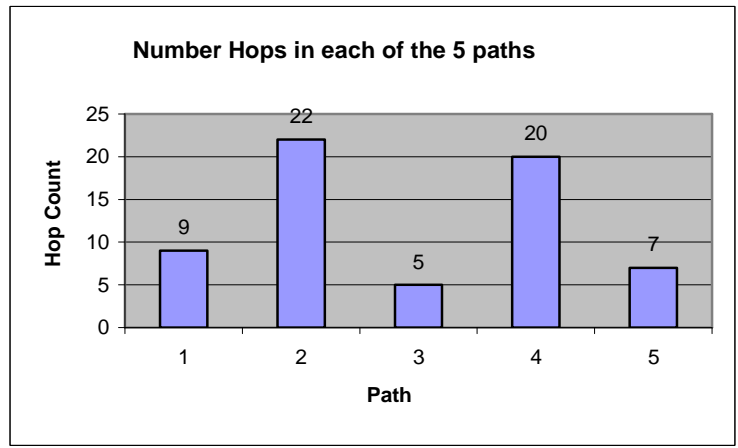

**Figure 1: Hop Count in different paths selected by Dijkstra's algorithm**

With the above parameters as input, the optimization algorithm divides data over each selected path. Figure 1 gives the number of hop counts in each path. Figure 2 gives the energy consumption in joules in the 3 Schemes i.e., when the data packets are routed via a single path (Scheme 1), when the data packets are equally distributed among 5 paths (Scheme 2) and when the data packets are routed using our proposed Scheme (Scheme 3). The energy comparisons are done with 100 and 200 data packets. We see that the energy consumption in our Scheme lies between that of single path and multipath equi-distribution Scheme but closer to the single path routing Scheme.

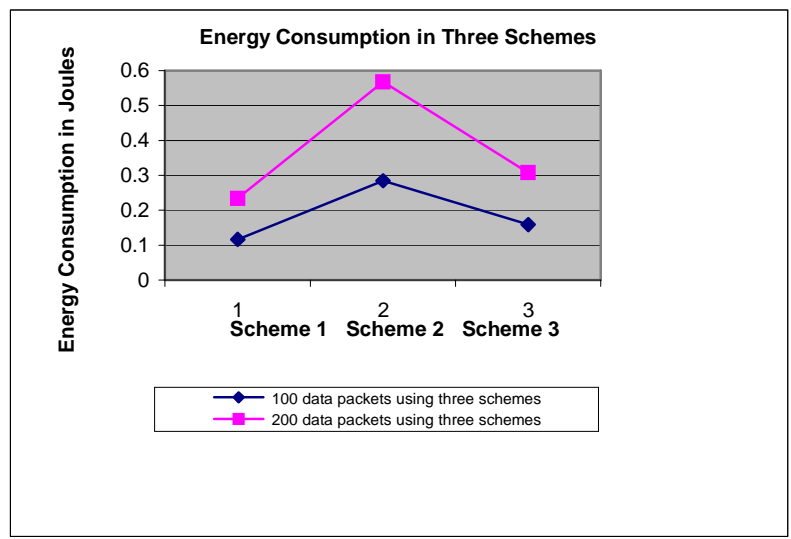

**Figure 2: Comparison of Energy Consumption in the 3 Schemes**

The Table 2 gives the data delivery time for individual paths using Scheme 2 and Scheme 3 for the 2 categories of simulation i.e., with 100 and 200 data packets.
Figure 3 compares the data delivery time in the three Schemes. The Scheme 1 causes maximum data delivery time while our proposed Scheme 3 gives minimum data delivery time. The data delivery time in Scheme 2 is in between the two. The data delivery time for Scheme 2 is less than than Scheme1 for using multiple paths. The data delivery time in Scheme 3 is even less than Scheme 2, as it takes various network factors in consideration during data distribution which Scheme 2 does not do. Figure 4 gives the data distribution in each path when simulation is done with 100 and 200 data packets in Scheme 3.

|       | Simulation using |          |                  |          |
|-------|------------------|----------|------------------|----------|
| Path  | 100 data packets |          | 200 data packets |          |
|       | Scheme 2         | Scheme 3 | Scheme 2         | Scheme 3 |
| 1     | 3.595            | 3.594    | 7.194            | 7.194    |
| 2     | 8.794            | 3.514    | 17.59            | 7.034    |
| 3     | 1.994            | 3.694    | 3.994            | 7.394    |
| 4     | 7.994            | 3.594    | 15.994           | 7.194    |
| 5     | 2.794            | 3.634    | 5.594            | 7.274    |

**Table 2: Delay in the individual path (in secs) in Scheme 2 and Scheme 3 when simulation is done with 100 and 200 data packets**

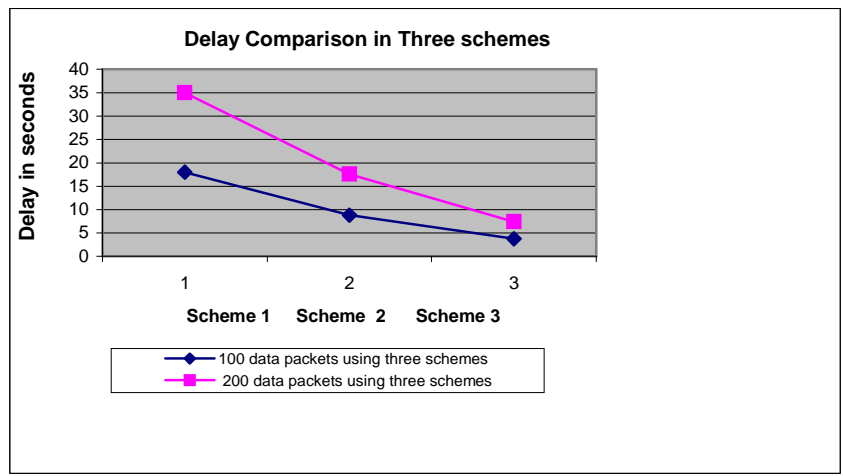

**Figure 3: Delay comparison in the three schemes**

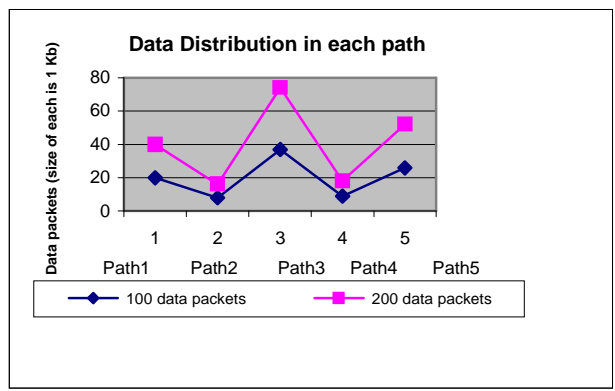

**Figure 4: Data Distribution in the five paths in Scheme 3**

From the above results we have observed that the multi-path routing is the best of the three Schemes where the net power consumption and the net end-to-end delay both are minimized.

### 5.2 Simulation results for the Multi-Source model

In order to evaluate the performance of the MADDR we have tested it against 3 frameworks in NS2. In each of the frameworks, we have pushed 1000 and 2000 data packets (size of each data packet being 1 Kb) from each of 3 sources to a common sink. Each source uses 3 locally node-disjoint paths to reach the destination. The paths of one source overlap with that of the other. In order to test the performance of the MADDR during heavy network load and strong overlap between the paths of different sources, as in a dense network, we have chosen a topology similar to the graph in Section 5.5. The figure given below (fig. 5) shows the number of hop counts in each of the 3 locally node-disjoint paths for each of the 3 Sources. In the simulation we make the 3 sources start together at the same time for the maximum contention.
In this simulation for the multi-source model in NS2, the value of sensing and processing energy ($K_r$) is taken as 81.2 µJ/sec. The value of the remaining network parameters is the same as in Table 1.

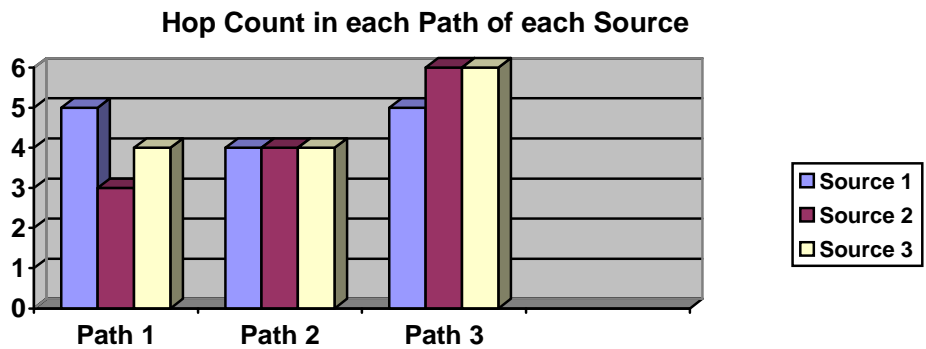

Figure 5 : Hop count in each of the 3 Paths for each of the 3 Sources

In the 1st framework, we have used traditional Mac 802.11 to send 1000 and 2000 data packets from each of the 3 sources to a common sink via 3 locally node-disjoint paths, sending the same data in each path. In the 2nd framework, we have used the MADDR over Mac 802.11 to transmit data to the sink using the same scenario as in framework 1, sending equal number of data packets in each of the 3 paths. Lastly in the 3rd framework, we have used the MADDR over Mac 802.11, but this time, to strategically distribute data packets among the 3 locally node-disjoint paths available using eqn [12]. Table 3 shows the number of data packets pushed in each of the 3 paths for each of the 3 sources in MADDR over Mac 802.11 using strategic Data Distribution, when each source transmits 1000 and 2000 data packets respectively.

|  | 1000 Data Packets | | | 2000 Data Packets | | |
| --- | --- | --- | --- | --- | --- | --- |
|  | Path 1 | Path 2 | Path 3 | Path 1 | Path 2 | Path 3 |
| Source 1 | 310 | 380 | 310 | 620 | 760 | 620 |
| Source 2 | 434 | 336 | 230 | 866 | 672 | 462 |
| Source 3 | 372 | 372 | 256 | 743 | 743 | 514 |

**Table 3 : Data packets pushed in each Path for each Source in MADDR over Mac 802.11**

**using strategic Data Distribution, each source transmitting 1000 and 2000 data packets**

Table 4 shows the net end-to-end data delivery time in each of the 3 frameworks when each source sends 1000 and 2000 data packets i.e. a total of 3000 and 6000 data packets are transmitted to a common sink. Table 5 shows the data delivery time (in seconds) for each source in each of the 3 frameworks.

|  | 1000 Data Packets Delay (in Secs) | 2000 Data Packets Delay (in Secs) |
|---|---|---|
| Traditional Mac 802.11 | 153 | 278 |
| MADDR over Mac 802.11 using Equal Data Distribution | 150 | 273 |
| MADDR over Mac 802.11 using Strategic Data Distribution | 139 | 262 |

**Table 4 : Net data delivery time (in seconds) in each framework**

|  | 1000 Data Packets | | | 2000 Data Packets | | |
|---|---|---|---|---|---|---|
|  | Source 1 | Source 2 | Source 3 | Source 1 | Source 2 | Source 3 |
| Traditional Mac 802.11 | 147 | 153 | 148 | 267 | 278 | 273 |
| MADDR over Mac 802.11 using Equal Data Distribution | 145 | 150 | 144 | 258 | 273 | 271 |
| MADDR over Mac 802.11 using Strategic Data Distribution | 139 | 138 | 139 | 256 | 255 | 262 |

**Table 5 : Data delivery time (in seconds) for each source in each framework**

Table 6 shows the net energy consumption (in Joules) in each of the 3 frameworks when each source sends 1000 and 2000 data packets i.e. a total of 3000 and 6000 data packets are transmitted to a common sink. Table 7 shows the energy consumption for each source in each of the 3 frameworks.

|  | 1000 Data Packets Energy Consumption (in Joules) | 2000 Data Packets Energy Consumption (in Joules) |
|---|---|---|
| Traditional Mac 802.11 | 0.237 | 0.466 |
| MADDR over Mac 802.11 using Equal Data Distribution | 0.234 | 0.463 |
| MADDR over Mac 802.11 using Strategic Data Distribution | 0.216 | 0.430 |

**Table 6: Net energy consumption (in Joules) in each framework**

|  | 1000 Data Packets | | | 2000 Data Packets | | |
|---|---|---|---|---|---|---|
|  | Source 1 | Source 2 | Source 3 | Source 1 | Source 2 | Source 3 |
| Traditional Mac 802.11 | 0.067 | 0.085 | 0.085 | 0.132 | 0.167 | 0.167 |
| MADDR over Mac 802.11 using Equal Data Distribution | 0.066 | 0.084 | 0.084 | 0.131 | 0.166 | 0.166 |
| MADDR over Mac 802.11 using Strategic Data Distribution | 0.061 | 0.078 | 0.077 | 0.122 | 0.155 | 0.153 |

**Table 7: Net energy consumption (in Joules) for each source in each framework**

### 5.3 Comparison of a few other routing protocols MADDR over 802.11

[11] gives an efficient multipath protocol (DCHT) for the wireless sensor network and establishes its efficiency over some other protocols like the Directed Diffusion [12],EDGE [13],C-MFR [14]. We have shown by simulation that the MADDR framework over 802.11 gives better throughput than DCHT over different network sizes. Hence from the above comparisons, it is apparent that MADDR over 802.11 performs better than the before mentioned protocols also. Furthermore, there are no packets losses in MADDR unlike the DCHT protocol where there are packet losses. Thus our suggested framework is more reliable. As the network size increases the throughput falls as the average path length increases.

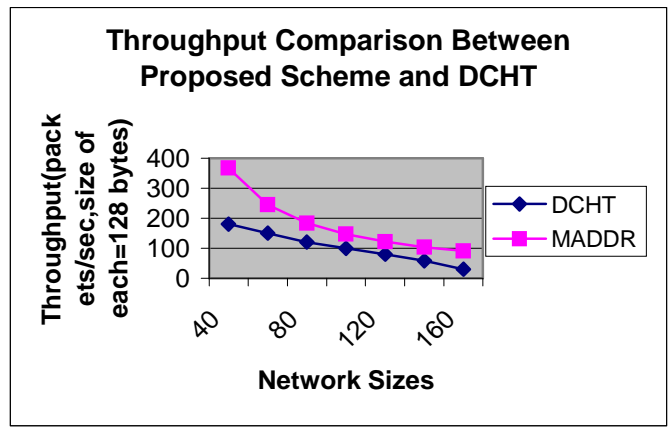

**Figure 6. Throughput of DCHT, Proposed Scheme with different network sizes**

### 6   CONCLUSION

From the above observations we can conclude that the MADDR serves well because it is adaptive and computes the data distribution based on various network factors. It has been observed from the simulations that the performance of the framework improves, with the increase in the amount of data and congestion in the network, over commonly used protocols. Also, the reduction in data delivery time and net energy consumption is significant as we have assumed a very low value for all the network parameters.

MADDR achieves its objective of both power and delay optimization. It is energy-aware, delay-aware and maintains an uniform load distribution. Uniformity and reliability is ensured by including nodes in a path based on a number of factors like hop-count, residual energy, bit rate and distributes data along the paths accordingly to make sure no node is overwhelmed with data it cannot handle. The efficient queue segmentation and data handling framework reduces the possibility of any particular link being over congested with data.

This framework can be implemented with any existing MAC protocol (it has been tested with 802.11,802.15.4) and improves its performance considerably. This proves to be one of the strongest features of the proposed framework.